\documentclass[review]{elsarticle}
\usepackage{amsfonts}
\usepackage{mathrsfs}
\usepackage{amssymb,amsmath}

\usepackage{algorithm}
\usepackage{algorithmic}
\usepackage{amsmath,amssymb,epsfig,graphics,subfigure}
\usepackage{theorem}
\usepackage{array,color}
\usepackage[compress]{cite}
\usepackage{amssymb}

\theoremheaderfont{\normalfont\bfseries}

\begin{document}
\begin{frontmatter}

\title{Cooperative Beamforming for Dual-Hop Amplify-and-Forward Multi-Antenna Relaying Cellular Networks}

\author[rvt]{Chengwen Xing\corref{cor1}}
 \ead{xingchengwen@gmail.com}
\author[focal]{Minghua Xia}
 \ead{minghua.xia@kaust.edu.sa}
\author[els]{Shaodan Ma}
 \ead{sdma@eee.hku.hk}
\author[els]{Yik-Chung Wu}
 \ead{ycwu@eee.hku.hk}
\cortext[cor1]{Corresponding author}

\address[rvt]{School of Information and Electronics, Beijing Institute of technology, Beijing,
China}
\address[focal]{King Abdullah University of Science and Technology (KAUST), Saudi Arabia}
\address[els]{Department of Electrical and Electronic Engineering,
The University of Hong Kong, Pokfulam Road, Hong Kong}


\begin{abstract}
In this paper, linear beamforming design for amplify-and-forward
relaying cellular networks is considered, in which
base station, relay station and mobile terminals are all equipped with multiple antennas.  The design is based on minimum mean-square-error criterion, and both uplink and downlink scenarios are considered. It is found that the downlink and uplink beamforming design problems are in the same form, and iterative algorithms with the same structure can be used to solve the design problems.  For the specific cases of fully loaded or overloaded uplink systems, a novel algorithm is derived and its relationships with several existing beamforming design algorithms for conventional MIMO or multiuser systems are revealed. Simulation results are presented to demonstrate the performance advantage of
the proposed design algorithms.
\end{abstract}


\end{frontmatter}

\section{Introduction}

\begin{figure}[!ht]
\centering
\includegraphics[width=.5\textwidth]{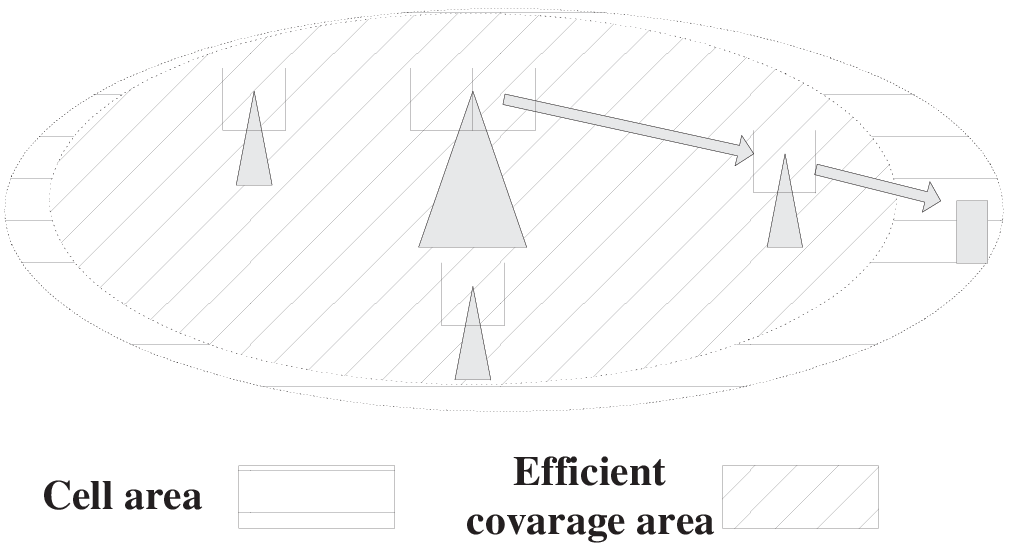}
\caption{Amplify-and-forward MIMO relaying cellular
network.}\label{fig:0}
\end{figure}

Cooperative communication is a promising technology to improve
quality and reliability of wireless links
\citep{Scaglione06,Laneman04,Guan08,Tang07,Rong09,Behbahani08,Bolcskei06,Wang05}.
One of the most important application scenarios of cooperative
communications is cellular network.  Due to shadowing or deep fading
of wireless channels, base station may not be able to sufficiently
cover all mobile terminals in a cell, especially those on the edge.
Deployment of relay stations is an effective and economic way to
improve the communication quality in cellular networks, as shown in
Fig. \ref{fig:0}.

In cooperative cellular networks, there are two major strategies in
relaying.  Relay station can either decode the received signal
before retransmission \citep{Chae08} or simply amplify-and-forward
(AF) the received signal to the corresponding destination without
decoding \citep{Zhang09}.  AF strategy has low complexity and
minimal processing delay, and is more secure. These reasons make AF
preferable in practical implementation.  In fact, deployment of AF
relay station with multiple antenna to enlarge coverage of base
station is one of the most important components in the future
communication protocols, e.g., LTE, IMT-Advanced and Winner project
\citep{Stefania09}, \citep{Osseiran09}.

With multiple antennas at mobile terminals, relay station and base
station, a natural question is how to allocate limited power
resource in the spatial domain.  In general, power allocation is
equivalent to beamforming matrices design at base station, relay and
mobile terminals, and the objective  can be maximizing capacity
\citep{Telatar99} or minimizing the mean-square error (MSE) of the
recovered data \citep{Kay93}.  The MSE criterion is a widely chosen
one since it aims at the data be recovered as accurate as possible,
and is extensively used in power allocation in classical
point-to-point \citep{Larsson03,Sampath01,Tse05} or multi-user MIMO
systems\citep{Zhang05,Shi08,Serbetli04,Christensen08,Codreanu07,Hunger09,Shi2007}.
The MSE minimization is also related to capacity maximization \citep
{Sampath01}, \citep{Christensen08} if a suitable weighting is
applied to different data streams.

In a cellular network, the base station and relay station are
usually allowed to be equipped with multiple antennas.  For each
mobile terminal, if it is equipped with single antenna, such relay
cellular networks has been investigated from various point-of-views.
For example, beamforming design for capacity maximization has been
considered in \citep{Chae08}, and quality-of-service based power
control has been investigated in \citep{Zhang09}. However, in the
next generation multi-media wireless communications, it is likely
that the size of a mobile terminal allows multiple antennas to be
deployed. Unfortunately, extension from the previous works on single
antenna mobile terminals to multi-antenna terminals is by no mean
straightforward.

In this paper, we take a step further  to consider the case where
each mobile terminal is also equipped with multiple antennas.  In
particular, we consider the joint precoders, forwarding matrix, and
equalizers design for both uplink and downlink AF relaying cellular
network, under power constraints. The design problems are
formulated as optimization problem minimizing the sum MSE of
multiple detected data streams. While extension of the presented algorithm to weighted MSE criterion is straightforward, we focus on sum MSE for notational clarity. The contribution of
the paper is as follows.  Firstly, in the downlink, the precoder at
base station, forwarding matrix at relay station and equalizers
at mobile terminals are jointly designed by an iterative algorithm.
Secondly, in the uplink case, we demonstrate that the formulation of
the beamforming design problem has the same form as that in the
downlink, and the same iterative algorithm can be employed.
Thirdly, since the general iterative solution provides little insight, we derive another algorithm
under the specific case when the number of independent data streams
from different mobile terminals is greater than or equal to their number of antenna.  It
is found that the resultant solution includes several existing
algorithms for multi-user MIMO or AF relay network with single
antenna as special cases.

The paper is organized as follows.  In Section~\ref{sect:downlink}, beamforming design
problem in downlink is investigated, and an iterative algorithm is
presented.  In Section~\ref{sect:uplink}, the analogy of the uplink and downlink beamforming design problems is demonstrated.  Furthermore, another beamforming design algorithm is derived for the specific case of fully loaded or overloaded system, and the relationships of this algorithm
with other existing algorithms are discussed.  Simulation results
are given in Section~\ref{sect:simul} to demonstrate the effectiveness of the
proposed algorithms.  Finally, conclusions are drawn in Section~\ref{sect:conclusion}.

The following notations are used throughout this paper. Boldface
lowercase letters denote vectors, while boldface uppercase letters
denote matrices. The notation ${\bf{Z}}^{\rm{H}}$ denotes the
Hermitian of the matrix ${\bf{Z}}$, and ${\rm{Tr}}({\bf{Z}})$ is the
trace of the matrix ${\bf{Z}}$. The symbol ${\bf{I}}_{M}$ denotes an
$M \times M$ identity matrix, while ${\bf{0}}_{M,N}$ denotes an $M
\times N$ all zero matrix. The notation ${\bf{Z}}^{1/2}$ is the
Hermitian square root of the positive semidefinite matrix
${\bf{Z}}$, such that ${\bf{Z}}^{1/2}{\bf{Z}}^{1/2}={\bf{Z}}$ and
${\bf{Z}}^{1/2}$ is also a Hermitian matrix. The operation
${\rm{diag}}\{[{\bf{A}} \ {\bf{B}}]\}$ is defined as a block
diagonal matrix with ${\bf{A}}$ and ${\bf{B}}$ as block diagonal.
The symbol ${\mathbb{E}}\{\bullet\}$ represents the statistical expectation.
The operation ${\rm{vec}}({\bf{Z}})$ stacks the columns of the
matrix ${\bf{Z}}$ into a single vector. The symbol $\otimes$ denotes
the Kronecker product. For two Hermitian matrices, ${\bf{C}} \succeq
{\bf{D}}$ means that ${\bf{C}}-{\bf{D}}$ is a positive semi-definite
matrix.

\section{Downlink Beamforming Design}
\label{sect:downlink}

\subsection{System model and problem formulation}

\begin{figure}[!ht]
\centering
\includegraphics[width=1.0\textwidth]{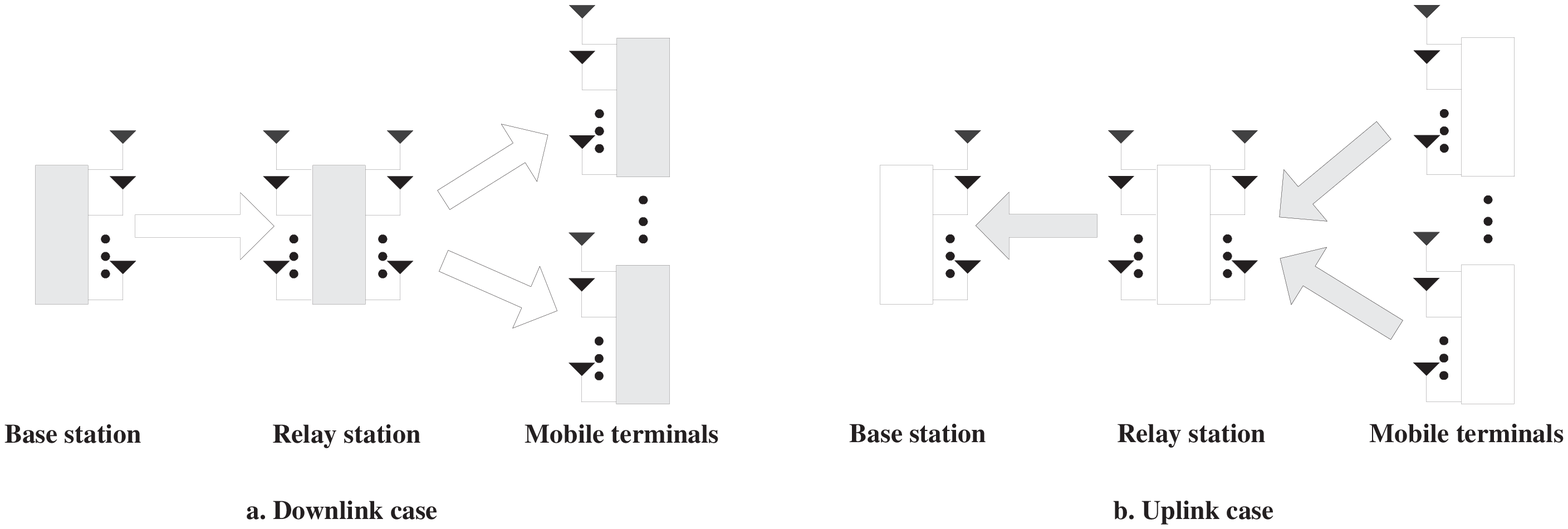}
\caption{Amplify-and-forward MIMO relaying downlink and uplink
cellular systems.}\label{fig:2}
\end{figure}

On the boundary of a cell, due to shadowing or deep fading, the
direct link between base station (BS) and mobile terminals may not
be good enough to maintain normal communication. Then mobile
terminals will rely on relay station to communicate with BS. As
shown in Fig.~\ref{fig:2}a, in downlink, signal is first transmitted
from the BS to the relay station and then the relay station forwards
the received signal to the corresponding mobile terminals. It is
assumed that the BS has $N_{B}$ antennas and the relay station has
$N_{R}$ antennas. For the $k^{\rm{th}}$ mobile terminal, it has
$N_{M,k}$ antennas. The BS needs to simultaneously
communicate with $K$ mobile terminals via a single relay station.
There are $L_k$ data streams to be transmitted from the BS to the
$k^{\rm{th}}$ mobile terminal, and the signal for the $k^{\rm{th}}$
mobile terminal is denoted by a $L_k \times 1$ vector ${\bf{s}}_k$.
It is assumed that different data streams are independent, i.e.,
$\mathbb{E}\{{\bf{s}}_k{\bf{s}}_j^{\rm{H}}\}={\bf{0}}_{L_k,L_j}$ when $k \ne
j$ and $\mathbb{E}\{{\bf{s}}_k{\bf{s}}_k^{\rm{H}}\}={\bf{I}}_{L_k}$.
With separate precoder ${\bf{T}}_k$ for different mobile terminals,
the received signal at the relay station is
\begin{align}
\label{D_R}
{\bf{r}}={\bf{H}}_{BR}{\bf{T}}{\bf{s}}+{\boldsymbol{\eta}},
\end{align}where ${\bf{H}}_{BR}$ denotes the $N_B \times N_R$ channel matrix between the BS and
relay station, $ {\bf{T}}=[{\bf{T}}_1, \ \cdots, \ {\bf{T}}_K]$,
${\bf{s}}=[{\bf{s}}_1^{\rm{T}}, \ \cdots, \
{\bf{s}}_K^{\rm{T}}]^{\rm{T}}$ and the vector ${\boldsymbol{\eta}}$
denotes the additive Gaussian noise with zero mean and covariance
matrix ${\bf{R}}_{{\boldsymbol{\eta}}}$. The power constraint at the
BS is given by $\sum_k{\rm{Tr}}({\bf{T}}_k{\bf{T}}_k^{\rm{H}}) \le
P_s$, where
$P_s$ is the maximum transmit power. 

At the relay station, before retransmission  the signal ${\bf{r}}$ is
multiplied with a forwarding matrix ${\bf{W}}$ under a power
constraint ${\rm{Tr}}({\bf{W}}{\bf{R}}_{{\bf{r}}}{\bf{W}})^{\rm{H}}
\le P_r$, where $P_r$ is the maximum transmit power at the relay
station and ${\bf{R}}_{{\bf{r}}}$ is the covariance matrix of the received signal ${\bf{r}}$:
\begin{align}
{\bf{R}}_{\bf{r}}&={\bf{H}}_{BR}{\bf{T}}{\bf{T}}^{\rm{H}}{\bf{H}}_{BR}^{\rm{H}}+{\bf{R}}_{{\boldsymbol{\eta}}}.
\end{align}Finally, at the $k^{\rm{th}}$ mobile terminal, the received signal
${\bf{y}}_k$ is
\begin{align}
{\bf{y}}_k={\bf{H}}_{RM,k}{\bf{W}}{\bf{H}}_{BR}
{\bf{T}}{\bf{s}}
+{\bf{H}}_{RM,k}{\bf{W}}{\boldsymbol{\eta}}+{\bf{v}}_{k},
\end{align} where matrix ${\bf{H}}_{RM,k}$ is the $N_R \times N_{M,k}$ channel matrix between the relay station and the $k^{\rm{th}}$ mobile
terminal, and ${\bf{v}}_{k}$ is the additive Gaussian noise at the $k^{\rm{th}}$ mobile terminal with zero mean and  covariance matrix ${\bf{R}}_{{\bf{v}}_{k}}$.
%

At each mobile terminal, an equalizer ${\bf{G}}_k$ is employed to
detect the data. The mean-square-error (MSE) of data detection at the  $k^{\rm{th}}$
terminal is
\begin{align}
\label{MSE_down_a}
&{\rm{MSE}}_k({\bf{G}}_k,{\bf{W}},{\bf{T}}_k)\\&={\mathbb{E}}\{\|{\bf{G}}_k{\bf{y}}_k-{\bf{s}}_k\|^2 \} \nonumber \\
&={\rm{Tr}}({\bf{G}}_k({\bf{H}}_{RM,k}{\bf{W}}{\bf{R}}_{{\bf{r}}}{\bf{W}}^{\rm{H}}
{\bf{H}}_{RM,k}^{\rm{H}}+{\bf{R}}_{{\bf{v}}_{k}}){\bf{G}}_k^{\rm{H}})-{\rm{Tr}}({\bf{G}}_k{\bf{H}}_{RM,k}{\bf{W}}{\bf{H}}_{BR}{\bf{T}}_k)\nonumber \\
&-{\rm{Tr}}(({\bf{G}}_k{\bf{H}}_{RM,k}{\bf{W}}{\bf{H}}_{BR}{\bf{T}}_k)^{\rm{H}})
+{\rm{Tr}}({\bf{I}}_{L_k}).
\end{align}
Now defining ${\bf{y}}=[{\bf{y}}_{1}^{\rm{T}}, \ \cdots, \
{\bf{y}}_{K}^{\rm{T}}]^{\rm{T}}$,
${\bf{H}}_{RM}=[{\bf{H}}_{RM,1}^{\rm{T}}, \ \cdots, \
{\bf{H}}_{RM,K}^{\rm{T}}]^{\rm{T}}$,
${\bf{v}}=[{\bf{v}}_{1}^{\rm{T}}, \ \cdots, \
{\bf{v}}_{K}^{\rm{T}}]^{\rm{T}}$, and
${\bf{G}}={\rm{diag}}\{[{\bf{G}}_1, \ \cdots, \ {\bf{G}}_K]\}$,
the sum MSE can be written as
\begin{align}
{\rm{{MSE}}}_{D}({\bf{G}},{\bf{W}},{\bf{T}})=& \sum_{k=1}^K{\rm{MSE}}_k({\bf{G}}_k,{\bf{W}},{\bf{T}}_k) \nonumber \\
=&{\rm{Tr}}({\bf{G}}({\bf{H}}_{RM}{\bf{W}}{\bf{R}}_{{\bf{r}}}{\bf{W}}^{\rm{H}}
{\bf{H}}_{RM}^{\rm{H}}+{\bf{R}}_{{\bf{v}}}){\bf{G}}^{\rm{H}})-{\rm{Tr}}({\bf{G}}{\bf{H}}_{RM}{\bf{W}}{\bf{H}}_{BR}{\bf{T}})\nonumber \\
&-{\rm{Tr}}(({\bf{G}}{\bf{H}}_{RM}{\bf{W}}{\bf{H}}_{BR}{\bf{T}})^{\rm{H}})
+{\rm{Tr}}({\bf{I}}_{L}),
\end{align} where $L=\sum_{k=1}^KL_{k}$ and ${\bf{R}}_{{\bf{v}}}={\rm{diag}}\{[{\bf{R}}_{{\bf{v}}_1}, \ \cdots, \
{\bf{R}}_{{\bf{v}}_K}]\}$.

Therefore, the downlink beamforming optimization problem can be formulated as
\begin{align}
\label{prob:opt_orig}
& {\min_{{\bf{G}},{\bf{W}},{\bf{T}}} } \ \ \ \ \ {\rm{{MSE}}}_{D}({\bf{G}},{\bf{W}},{\bf{T}}) \nonumber \\
&  \ \ {\rm{s.t.}} \ \ \ \ \ \  {\rm{Tr}}({\bf{T}}{\bf{T}}^{\rm{H}}) \le P_{s} \nonumber \\
& \ \ \ \ \ \ \ \ \ \ \ \
{\rm{Tr}}({\bf{W}}{\bf{R}}_{{\bf{r}}}{\bf{W}}^{\rm{H}}) \le P_{r}\nonumber \\
& \ \ \ \ \ \ \ \ \ \ \ \ {\bf{G}}={\rm{diag}}\{[{\bf{G}}_1, \ \cdots, \ {\bf{G}}_K]\}.
\end{align}

The optimization problem (\ref{prob:opt_orig}) is a nonconvex
optimization problem for ${\bf{T}}$, ${\bf{W}}$ and ${\bf{G}}$, and
there is no closed-form solution. This challenge remains even for
the special case of multiuser MIMO systems \citep{Zhang05},
\citep{Shi08}, \citep{Hunger09} where only single hop transmission
is involved. However, notice that when two out of the three
variables are fixed, the optimization problem (\ref{prob:opt_orig})
for the remaining variable is a convex problem, and thus can be
solved. Therefore, an iterative algorithm alternating the design of
three variables can be employed.

\subsection{Proposed iterative algorithm}

\noindent (1) \textbf{\underline{Equalizer design at the
destination}}

When ${\bf{T}}$ and ${\bf{W}}$ are fixed, the optimization
problem (\ref{prob:opt_orig}) is an unconstrained convex quadratic
optimization problem for ${\bf{G}}$. Furthermore, since the structure of ${\bf{G}}$ is block diagonal, the design of individual ${\bf{G}}_k$ are decoupled.
Therefore, the necessary and sufficient condition for the optimal
solution is
\begin{align}
\frac{\partial \sum_k
{\rm{MSE}}_k({\bf{G}}_k,{\bf{W}},{\bf{T}}_k)}{\partial
{\bf{G}}_k^{*}}={\bf{0}}_{L_k,N_{M,k}},
\end{align} and the optimal equalizer for the $k^{\rm{th}}$ mobile terminal can be easily shown to be
\begin{align}
\label{G_equa} &{\bf{G}}_k
=({\bf{H}}_{RM,k}{\bf{W}}{\bf{H}}_{BR}{\bf{T}}_k)^{\rm{H}}({\bf{H}}_{RM,k}
{\bf{W}}{\bf{R}}_{{\bf{r}}}{\bf{W}}^{\rm{H}}{\bf{H}}_{RM,k}^{\rm{H}}
+{\bf{R}}_{{\bf{v}}_{k}})^{-1}.
\end{align}

\noindent (2) \textbf{\underline{Forwarding matrix design at the
relay station}}

When ${\bf{T}}$ and ${\bf{G}}$ are fixed, the optimization problem
(\ref{prob:opt_orig}) is a constrained convex optimization problem
for the variable ${\bf{W}}$, and the Karush-Kuhn-Tucker (KKT)
conditions are the necessary and sufficient conditions for the
optimal solution \citep{Beck07}. The KKT conditions of the
optimization problem (\ref{prob:opt_orig}) with respective to
${\bf{W}}$ are \citep{Boyd04}
\begin{align}
&{\bf{H}}_{RM}^{\rm{H}}{\bf{G}}^{\rm{H}}{\bf{G}}{\bf{H}}_{RM}{\bf{W}}
{\bf{R}}_{{\bf{r}}}+\lambda{\bf{W}} {\bf{R}}_{{\bf{r}}}
=({\bf{H}}_{BR}{\bf{T}}{\bf{G}}{\bf{H}}_{RM})^{\rm{H}} \label{KKT_1} \\
&\lambda({\rm{Tr}}({\bf{W}}{\bf{R}}_{{\bf{r}}}{\bf{W}}^{\rm{H}})-P_{r})=0, \ \ \ \lambda \ge 0, \label{KKT_2} \\
& {\rm{Tr}}({\bf{W}}{\bf{R}}_{{\bf{r}}}{\bf{W}}^{\rm{H}}) \le
P_{r},\label{KKT_3}
\end{align} where $\lambda$ is the Lagrange multiplier.

Based on the first KKT condition (\ref{KKT_1}), the optimal
forwarding matrix ${\bf{W}}$ can be written as
\begin{align}
\label{F_equa}
{\bf{W}}&=({\bf{H}}_{RM}^{\rm{H}}{\bf{G}}^{\rm{H}}{\bf{G}}{\bf{H}}_{RM}
+\lambda{\bf{I}})^{-1}({\bf{H}}_{BR}{\bf{T}}{\bf{G}}{\bf{H}}_{RM})^{\rm{H}}
{\bf{R}}_{{\bf{r}}}^{-1},
\end{align} where the value of $\lambda$ is computed using
(\ref{KKT_2}) and (\ref{KKT_3}). Since $\lambda$ also appears in ${\bf{W}}$, (\ref{KKT_2}) and (\ref{KKT_3}) depends on $\lambda$ in a nonlinear way and there is no closed-form solution.  Below, we propose a low complexity method to solve
(\ref{KKT_2}) and (\ref{KKT_3}).

First, notice that in order to have (\ref{KKT_2}) satisfied, either
$\lambda=0$ or
${\rm{Tr}}({\bf{W}}{\bf{R}}_{{\bf{r}}}{\bf{W}}^{\rm{H}})=P_{r}$ must
hold. If $\lambda=0$ also makes (\ref{KKT_3}) satisfied, $\lambda=0$
is a solution to (\ref{KKT_2}) and (\ref{KKT_3}). On other hand, if
$\lambda=0$ does not make (\ref{KKT_3}) satisfied, we have to solve
${\rm{Tr}}({\bf{W}}{\bf{R}}_{{\bf{r}}}{\bf{W}}^{\rm{H}})=P_{r}$. It
can be proved that \citep{Xing10} when  ${\bf{T}}$ and ${\bf{G}}$
are fixed, the function $
f(\lambda)={\rm{Tr}}({\bf{W}}{\bf{R}}_{{\bf{r}}}{\bf{W}}^{\rm{H}})$
is a decreasing function of ${\lambda}$ and the range of ${\lambda}$
must be within
\begin{align}
 0 \le \lambda \le \sqrt{\frac{{\rm{Tr}}({\bf{E}}{\bf{R}}_{{\bf{r}}}^{-1}{\bf{E}}^{\rm{H}})}{P_r}}
\end{align} where
${\bf{E}}=\sum_k\{({\bf{H}}_{BR}{\bf{T}}_k{\bf{G}}_k{\bf{H}}_{RM,k})^{\rm{H}}\}$.
Therefore, $\lambda$ can be efficiently computed by one-dimension
search, such as bisection search or golden search. Since
${\rm{Tr}}({\bf{W}}{\bf{R}}_{\bf{r}}{\bf{W}}^{\rm{H}})=P_{r}$ is a
stronger condition than
${\rm{Tr}}({\bf{W}}{\bf{R}}_{\bf{r}}{\bf{W}}^{\rm{H}})\le P_{r}$,
(\ref{KKT_3}) is satisfied automatically in this case. In summary, $\lambda$ is computed as
\begin{equation}
\label{gamma}
\lambda=\begin{cases} 0 & \text{if $f(0) \le P_r$} \\
\text{Solve $f(\lambda)=P_r$ using bisection algorithm} &
\text{Otherwise}
\end{cases}.
\end{equation}

\noindent (3) \textbf{\underline{Precoder design at the BS}}

When ${\bf{W}}$ and ${\bf{G}}$ are fixed, the optimization problem
(\ref{prob:opt_orig}) can be straightforwardly formulated as the following convex
quadratic optimization problem for the precoder ${\bf{T}}$
\begin{align}
\label{Prob:Qua}
& {\min_{{\bf{T}}}} \ \ \ {\rm{Tr}}({\bf{N}}_0^{\rm{H}}{\bf{T}}^{\rm{H}}{\bf{A}}_0{\bf{T}}{\bf{N}}_0)+2\mathcal{R}\{{\rm{Tr}}({\bf{B}}_0^{\rm{H}}{\bf{T}})\}+c_0 \nonumber \\
& {\rm{s.t.}} \ \ \ \
{\rm{Tr}}({\bf{N}}_1^{\rm{H}}{\bf{T}}^{\rm{H}}{\bf{A}}_1{\bf{T}}{\bf{N}}_1)
+2\mathcal{R}\{{\rm{Tr}}({\bf{B}}_1^{\rm{H}}{\bf{T}})\}+c_1\le
0,\nonumber \\
& \ \ \ \ \ \ \ \
{\rm{Tr}}({\bf{N}}_2^{\rm{H}}{\bf{T}}^{\rm{H}}{\bf{A}}_2{\bf{T}}{\bf{N}}_2)+2\mathcal{R}\{{\rm{Tr}}
({\bf{B}}_2^{\rm{H}}{\bf{T}})\}+c_2\le 0,
\end{align} where the corresponding parameters are defined as
\begin{align}
{\bf{A}}_0&={\bf{H}}_{BR}^{\rm{H}}{\bf{W}}^{\rm{H}}{\bf{H}}_{RM}^{\rm{H}}{\bf{G}}^{\rm{H}}
{\bf{G}}{\bf{H}}_{RM}{\bf{W}}{\bf{H}}_{BR}, \ \ \
{\bf{A}}_1={\bf{I}}, \ \ \ {\bf{A}}_2={\bf{H}}_{BR}^{\rm{H}}{\bf{W}}^{\rm{H}}{\bf{W}}{\bf{H}}_{BR}, \nonumber \\
{\bf{B}}_0^{\rm{H}}&=-{\bf{G}}{\bf{H}}_{RM}{\bf{W}}{\bf{H}}_{BR}, \
\ \ \ \ \ \ \ \ \ \ \ \ \ \ \ \ \ \ \ \ \  {\bf{B}}_1={\bf{B}}_2={\bf{0}}, \nonumber \\
{\bf{N}}_0& ={\bf{N}}_1={\bf{N}}_2={\bf{I}}_L, \nonumber\\
c_0&={\rm{Tr}}({\bf{R}}_{{\boldsymbol{\eta}}}{\bf{W}}^{\rm{H}}{\bf{H}}_{RM}^{\rm{H}}{\bf{G}}^{\rm{H}}
{\bf{G}}{\bf{H}}_{RM}{\bf{W}}) +{\rm{Tr}}({\bf{I}}_{L})+{\rm{Tr}}({\bf{G}}{\bf{R}}_{{\bf{v}}}{\bf{G}}^{\rm{H}})) \nonumber \\
c_1&=-P_s, \ \ \
c_2={\rm{Tr}}({\bf{W}}{\bf{R}}_{{\boldsymbol{\eta}}}{\bf{W}}^{\rm{H}})-P_{r}.
\end{align}Notice that the objective function and the constraints are of the same form. Using the property ${\rm{Tr}}({\bf{A}}{\bf{B}})={\rm{vec}}^{\rm{H}}({\bf{A}}^{\rm{H}}){\rm{vec}}({\bf{B}})$ and the property of Kronecker product,
we can write $(l=0,1,2)$
\begin{align} {\rm{Tr}}({\bf{N}}_l^{\rm{H}}{\bf{T}}^{\rm{H}}{\bf{A}}_l{\bf{T}}{\bf{N}}_l)
=&{\rm{Tr}}({\bf{N}}_l^{\rm{H}}{\bf{T}}^{\rm{H}}{\bf{A}}_l^{\frac{\rm{H}}{2}}{\bf{A}}_l^{\frac{1}{2}}
{\bf{T}}{\bf{N}}_l)
\nonumber \\
=&{\rm{vec}}^{\rm{H}}({\bf{A}}_l^{\frac{1}{2}}{\bf{T}}{\bf{N}}_l){\rm{vec}}({\bf{A}}_l^{\frac{1}{2}}{\bf{T}}{\bf{N}}_l)
\nonumber \\
=&{\rm{vec}}^{\rm{H}}({\bf{T}})({\bf{N}}_l^{*}\otimes
{\bf{A}}_l^{\frac{\rm{H}}{2}}) ({\bf{N}}_l^{\rm{T}}\otimes
{\bf{A}}_l^{\frac{1}{2}}){\rm{vec}}({\bf{T}}),
\end{align} where the first equality is based on the fact that
${\bf{A}}_{l}$'s are positive semidefinite  matrices. Furthermore,
we can also write
${\rm{Tr}}({\bf{B}}_l^{\rm{H}}{\bf{T}})={\rm{vec}}^{\rm{H}}({\bf{B}}_l^{\rm{H}}){\rm{vec}}({\bf{T}})$.
Putting these two results into (\ref{Prob:Qua}) and after
introducing an auxiliary variable $t$ \citep{Vandenberghe96},
(\ref{Prob:Qua}) is equivalent to the following optimization problem
\begin{align}
\label{Prob:Qua_a}
& {\min_{{\bf{T}},t}} \ \ \ t \nonumber \\
& {\rm{s.t.}} \ \ \ \
{\rm{vec}}^{\rm{H}}({\bf{T}})({\bf{N}}_0^{*}\otimes
{\bf{A}}_0^{\frac{\rm{H}}{2}}) ({\bf{N}}_0^{\rm{T}}\otimes
{\bf{A}}_0^{\frac{1}{2}}){\rm{vec}}({\bf{T}}) \le
t-2\mathcal{R}\{{\rm{vec}}^{\rm{H}}({\bf{B}}_0^{\rm{H}}){\rm{vec}}({\bf{T}})\} \nonumber \\
& \ \ \ \ \ \ \ \
{\rm{vec}}^{\rm{H}}({\bf{T}})({\bf{N}}_1^{*}\otimes
{\bf{A}}_1^{\frac{\rm{H}}{2}}) ({\bf{N}}_1^{\rm{T}}\otimes
{\bf{A}}_1^{\frac{1}{2}}){\rm{vec}}({\bf{T}}) \le
-c_1-2\mathcal{R}\{{\rm{vec}}^{\rm{H}}({\bf{B}}_1^{\rm{H}}){\rm{vec}}({\bf{T}})\}
\nonumber \\
& \ \ \ \ \ \ \ \
{\rm{vec}}^{\rm{H}}({\bf{T}})({\bf{N}}_2^{*}\otimes
{\bf{A}}_2^{\frac{\rm{H}}{2}}) ({\bf{N}}_2^{\rm{T}}\otimes
{\bf{A}}_2^{\frac{1}{2}}){\rm{vec}}({\bf{T}}) \le
-c_2-2\mathcal{R}\{{\rm{vec}}^{\rm{H}}({\bf{B}}_2^{\rm{H}}){\rm{vec}}({\bf{T}})\}.
\end{align} Since $c_0$ does not affect the optimization problem, it has been neglected in
 (\ref{Prob:Qua_a}).

With the Schur complement lemma \citep{Horn85}, the optimization
problem (\ref{Prob:Qua_a}) can be further reformulated as the
following semi-definite programming (SDP) problem
\citep{Vandenberghe96}
\begin{align}
\label{SDP}
&{\min_{{\bf{T}},t}} \ \ \ \ t \nonumber \\
&{\rm{s.t.}}  \ \ \left[ {\begin{array}{*{20}c}
   {\bf{I}} & ({\bf{N}}_0^{\rm{T}}\otimes {\bf{A}}_0^{\frac{1}{2}}){\rm{vec}}({\bf{T}})   \\
   (({\bf{N}}_0^{\rm{T}}\otimes {\bf{A}}_0^{\frac{1}{2}}){\rm{vec}}({\bf{T}}))^{\rm{H}} & -2\mathcal{R}\{{\rm{vec}}^{\rm{H}}({\bf{B}}_0){\rm{vec}}({\bf{T}})\}+t  \\
\end{array}} \right] \succeq 0 \nonumber  \\
&\ \ \ \ \ \ \left[ {\begin{array}{*{20}c}
   {\bf{I}} & ({\bf{N}}_l^{\rm{T}}\otimes {\bf{A}}_l^{\frac{1}{2}}){\rm{vec}}({\bf{T}})   \\
   (({\bf{N}}_l^{\rm{T}} \otimes {\bf{A}}_l^{\frac{1}{2}}){\rm{vec}}({\bf{T}}))^{\rm{H}} & -2\mathcal{R}\{{\rm{vec}}^{\rm{H}}({\bf{B}}_l){\rm{vec}}({\bf{T}})\}-c_l \\
\end{array}} \right]  \succeq 0, \ \ \ l=1,2.
\end{align}
The precoder at the BS is designed by solving this SDP problem using
standard numerical algorithms such as interior-point polynomial
algorithms \citep{Boyd04}, \citep{Vandenberghe96}.

\subsection{Summary and Initialization}

In summary, the downlink beamforming matrices are computed
iteratively. Since in each iteration, the MSE monotonically
decreases, the iterative algorithm is guaranteed to converge to at
least a local optimum. For initialization, identity matrices can be
chosen as initial values due to its simplicity and better
performance compared to randomly generated initial matrices
\citep{Zhang05}, \citep{Shi08}, \citep{Shi2007}. On the other hand,
we can also use a suboptimal design by viewing the downlink dual-hop
AF MIMO relay cellular networks as a combination of conventional
point-to-point MIMO system in the first hop, and multiuser MIMO
downlink system in the second hop. More specifically, for the first
hop, the linear minimum mean-square-error (LMMSE) precoder
${\bf{T}}$ at BS and equalizer ${\bf{W}}_{1}$ at relay station can
be jointly designed using the point-to-point water-filling solution
given in \citep{Sampath01}. For the second hop, the precoder
${\bf{W}}_{2}$ at relay station and equalizer ${\bf{G}}$ at mobile
terminals can be designed using the beamforming algorithm for
multiuser MIMO systems proposed in \citep{Wang05}. Based on the
results of ${\bf{W}}_{1}$ and ${\bf{W}}_{2}$, the forwarding matrix
at relay station equals to ${\bf{W}}={\bf{W}}_{1}{\bf{W}}_{2}$. We
refer this suboptimal algorithm as `separate LMMSE transceiver
design'. It will be shown in Simulation section that the convergence
speed using the second initialization is better than that of the
first one. Finally, the iterative design procedure is formally given
by

\noindent \underline{Algorithm 1}

\noindent With initial ${\bf{G}}^0$, ${\bf{W}}^0$ and ${\bf{T}}^0$, the algorithm proceeds iteratively and in each iteration:

\noindent (1) ${\bf{G}}$ is updated using (\ref{G_equa});

\noindent (2) ${\bf{W}}$ is updated using (\ref{F_equa}) and
(\ref{gamma});

\noindent (3) ${\bf{T}}$ is updated by solving (\ref{SDP}).

\noindent The algorithm stops when
$\|{\rm{MSE}}_D^{I}-{\rm{MSE}}_D^{I+1}\|\le {\mathcal{T}}_D$, where
${\rm{MSE}}_D^{I}$ is the total MSE in the $I^{\rm{th}}$ iteration
and $\mathcal{T}_D$ is a threshold value.


\section{Uplink Beamforming Design}
\label{sect:uplink}

\subsection{System model and analogy with downlink design}

In this section we will focus on beamforming matrices design
for uplink, as shown in Fig.~\ref{fig:2}b. In uplink, there are $L_k$ data streams to be transmitted from the $k^{\rm{th}}$ mobile terminal to the BS, and the signal from the $k^{\rm{th}}$ mobile terminal is denoted as ${\bf{s}}_k$. Without loss of generality, it is assumed that the
transmitted data streams are independent:
${\mathbb{E}}\{{\bf{s}}_k{\bf{s}}_j^{\rm{H}}\}={\bf{0}}_{L_k,L_j}$ when $k \ne j$ and
${\mathbb{E}}\{{\bf{s}}_k{\bf{s}}_k^{\rm{H}}\}={\bf{I}}_{L_k}$. At the $k^{\rm{th}}$ mobile terminal,
the transmit signal ${\bf{s}}_k$ is multiplied by a precoder matrix
${\bf{P}}_k$ under a power constraint
${\rm{Tr}}({\bf{P}}_k{\bf{P}}_k^{\rm{H}})\le P_{s,k}$, where
$P_{s,k}$ is the maximum transmit power at the $k^{\rm{th}}$ mobile
terminal. The received signal ${\bf{x}}$ at the relay station is the superposition of signals from different terminals through different channels and is given by
\begin{align}
\label{x_uplink}
{\bf{x}}={\bf{H}}_{MR}{\bf{P}}{\bf{s}}+{\bf{n}},
\end{align}where $
{\bf{H}}_{MR}\triangleq [{\bf{H}}_{MR,1} \ \cdots \ {\bf{H}}_{MR,K}
]$, $ {\bf{P}}\triangleq {\rm{diag}}\{[{\bf{P}}_1, \cdots,
{\bf{P}}_K]\}$, ${\bf{s}}\triangleq [{\bf{s}}_{1}^{\rm{T}} \
\cdots \ {\bf{s}}_{K}^{\rm{T}}]^{\rm{T}}$,  with
${\bf{H}}_{MR,k}$ being the $N_R \times N_{M,k}$ channel matrix between
the $k^{\rm{th}}$ mobile terminal and relay station, and
${{\bf{n}}}$ is the additive Gaussian noise at the relay station
with zero mean and covariance matrix ${\bf{R}}_{{\bf{n}}}$. Since
the data transmitted from different mobile terminals are independent,
the correlation matrix of ${\bf{x}}$ equals to
\begin{align}
\label{R_x_uplink}
{\bf{R}}_{{\bf{x}}}={\bf{H}}_{MR}{\bf{P}}{\bf{P}}^{\rm{H}}{\bf{H}}_{MR}
+{\bf{R}}_{{\bf{n}}}.
\end{align}

At the relay station, the received signal ${\bf{x}}$ is
multiplied with a linear forwarding matrix ${\bf{F}}$, with a power constraint $
{\rm{Tr}}({\bf{F}}{\bf{R}}_{\bf{x}}{\bf{F}}^{\rm{H}}) \le P_r$,
where $P_r$ is the maximum transmit power at the relay station.
Finally, the received signal at the BS is
\begin{align}
{\bf{y}}={\bf{H}}_{RB}{\bf{F}}{\bf{H}}_{MR}{\bf{P}}{\bf{s}}+{\bf{H}}_{RB}{\bf{F}}{\bf{n}}+{\boldsymbol{\xi}},
\end{align} where ${\bf{H}}_{RB}$ is the $N_B \times N_R$ channel matrix between the relay station and BS, and ${\boldsymbol{\xi}}$ is the additive zero mean Gaussian noise with covariance ${\bf{R}}_{{\boldsymbol \xi}}$.

When a linear equalizer ${\bf{B}}$ is adopted at the BS, the total MSE
of the detected data is
\begin{align}
\label{MSE_Orig}
{\rm{MSE}}_{U}({\bf{B}},{\bf{F}},{\bf{P}})
=& \mathbb{E}\{\|{\bf{B}}{\bf{y}}-{\bf{s}}\|^2\} \nonumber\\
=&
{\rm{Tr}}({\bf{B}}({\bf{H}}_{RB}{\bf{F}}{\bf{R}}_{{\bf{x}}}{\bf{F}}^{\rm{H}}{\bf{H}}_{RB}^{\rm{H}}+{\bf{R}}_{{\boldsymbol{\xi}}}){\bf{B}}^{\rm{H}})
-{\rm{Tr}}({\bf{B}}{\bf{H}}_{RB}{\bf{F}}{\bf{H}}_{MR}{\bf{P}})\nonumber \\
&-{\rm{Tr}}(({\bf{B}}{\bf{H}}_{RB}{\bf{F}}{\bf{H}}_{MR}{\bf{P}})^{\rm{H}})+{\rm{Tr}}({\bf{I}}_L),
\end{align}where $L=\sum_{k=1}^KL_k$ is the total number of data streams.
Finally, the optimization problem for beamforming matrices design
in the uplink case is formulated as
\begin{align}
\label{prob:uplink}
& \min_{{\bf{B}},{\bf{F}},{\bf{P}}} \ \ \ \ \ {\rm{MSE}}_U({\bf{B}},{\bf{F}},{\bf{P}}) \nonumber \\
& \ {\rm{s.t.}} \ \ \ \ \ \ \ {\rm{Tr}}({\bf{P}}_k{\bf{P}}_k^{\rm{H}}) \le P_{s,k}, \ \ k=1,\cdots,K \nonumber \\
& \ \ \ \ \ \ \ \ \ \ \ \ {\rm{Tr}}({\bf{F}}{\bf{R}}_{\bf{x}}{\bf{F}}^{\rm{H}})
\le P_{r} \nonumber \\
& \ \ \ \ \ \ \ \ \ \ \ \  {\bf{P}}={\rm{diag}}\{[{\bf{P}}_1,\ \cdots, \ {\bf{P}}_K]\}.
\end{align}
Comparing (\ref{prob:uplink}) with the downlink problem (\ref{prob:opt_orig}), it can be seen
that the two problems are in the same form, except that i) there are
individual constraints on ${\bf{P}}_k$ in (\ref{prob:uplink})
instead of a sum constraint on the corresponding ${\bf{T}}_k$ in
(\ref{prob:opt_orig}), and ii) the diagonal structure constraint is on precoder
instead of equalizer. However we can still employ the iterative
algorithm developed in the previous section for this uplink beamforming design problem. More
specifically, for equalizer ${\bf{B}}$ design, the problem is an
unconstrained convex optimization problem and the optimal solution
can be directly computed from the derivative of the objective
function. For forwarding matrix ${\bf{F}}$ design, the problem is a
convex quadratic optimization problem with only one constraint. In
this case, the optimal solution can be solved based on KKT
conditions. Finally,
for precoder ${\bf{P}}$ design, the problem is
 a convex quadratic optimization with multiple constraints, which can be transformed into a
 standard SDP problem. Notice that a SDP problem can handle any number of linear matrix inequality
 constraints and the diagonal structure of ${\bf{P}}$ does not
 affect the SDP problem.

%
%

Although the optimization  problem (\ref{prob:uplink}) can be solved
using an iterative algorithm alternating the three variables
${\bf{B}}$, ${\bf{F}}$ and ${\bf{P}}$, this solution provide little
insight into the nature of the problem. Below we consider the fully
loaded or overloaded MIMO systems in which the number of independent
data streams from mobile terminals is greater than or equal to the
number of its antennas, i.e., $N_{M,k}\le L_k$ \citep{Wong07},
\citep{Miguel10}. The solution is found to be insightful and
includes several existing algorithms for conventional AF MIMO relay
or multiuser MIMO as special cases.
%

\subsection{Uplink beamforming design for fully loaded or overloaded systems}

First, we reduce the number of variables of the optimization
problem. Noticing that there is no constraint on ${\bf{B}}$, the
optimal ${\bf{B}}$ satisfies ${\partial
{\rm{MSE}}_U({\bf{B}},{\bf{F}},{\bf{P}})}/{\partial
{{\bf{B}}}^{*}}={\bf{0}}_{L,N_B}$,  and the optimal equalizer at the
BS can be written as a function of forwarding matrix and precoder matrix. Therefore $
{\bf{B}}=({\bf{H}}_{RB}{\bf{F}}{\bf{H}}_{MR}{\bf{P}})^{\rm{H}}({\bf{H}}_{RB}{\bf{F}}
{\bf{R}}_{{\bf{x}}}{\bf{F}}^{\rm{H}}{\bf{H}}_{RB}^{\rm{H}}+{\bf{R}}_{{\boldsymbol{\xi}}})^{-1}$.
Substituting this result into (\ref{MSE_Orig}), the uplink MSE is
simplified as
\begin{align}
\label{MSE_uplink}
&{\rm{MSE}}_{U}({\bf{F}},{\bf{P}})\nonumber
\\
& ={\rm{Tr}}({\bf{I}}_L)-{\rm{Tr}}(({\bf{H}}_{RB}{\bf{F}}
{\bf{H}}_{MR}{\bf{P}})^{\rm{H}}({\bf{H}}_{RB}{\bf{F}}
{\bf{R}}_{{\bf{x}}}{\bf{F}}^{\rm{H}}{\bf{H}}_{RB}^{\rm{H}}+{\bf{R}}_{{\boldsymbol{\xi}}})^{-1}
({\bf{H}}_{RB}{\bf{F}}{\bf{H}}_{MR}{\bf{P}})).
\end{align}

Based on the definition of ${\bf{R}}_{\bf{x}}=
{\bf{H}}_{MR}{\bf{P}}{\bf{P}}^{\rm{H}}{\bf{H}}_{MR}^{\rm{H}}
+{\bf{R}}_{\bf{n}}$, it can be expressed as
\begin{align}
{\bf{R}}_{\bf{x}}
&={\bf{R}}_{{\bf{n}}}^{1/2}(\underbrace{{\bf{R}}_{{\bf{n}}}^{-1/2}
{\bf{H}}_{MR}{\bf{P}}{\bf{P}}^{\rm{H}}{\bf{H}}_{MR}^{\rm{H}}
{\bf{R}}_{{\bf{n}}}^{-1/2}+{\bf{I}}}_{\triangleq {\boldsymbol
\Xi}}){\bf{R}}_{{\bf{n}}}^{1/2}.
\end{align}
Now introducing ${\bf{\tilde
F}}={\bf{F}}{\bf{R}}_{\bf{n}}^{1/2}{\boldsymbol \Xi}^{1/2}$, the MSE
(\ref{MSE_uplink}) becomes
\begin{align}
\label{MSE_uplink_a} {\rm{\overline{MSE}}}_{U}({\bf{\tilde
F}},{\bf{P}})
 =&{\rm{Tr}}({\bf{I}}_L)-{\rm{Tr}}(({\bf{H}}_{RB}{\bf{\tilde
F}}{\boldsymbol{\Xi}}^{-1/2}{\bf{R}}_{{\bf{n}}}^{-1/2}
{\bf{H}}_{MR}{\bf{P}})^{\rm{H}}\nonumber \\
& \times ({\bf{H}}_{RB}{\bf{\tilde F}}{\bf{\tilde
F}}^{\rm{H}}{\bf{H}}_{RB}^{\rm{H}}+{\bf{R}}_{{\boldsymbol{\xi}}})^{-1}({\bf{H}}_{RB}{\bf{\tilde
F}}{\boldsymbol{\Xi}}^{-1/2}{\bf{R}}_{{\bf{n}}}^{-1/2}{\bf{H}}_{MR}{\bf{P}})).
\end{align}
Thus the uplink beamforming design optimization problem
(\ref{prob:uplink}) is rewritten as
\begin{align}
\label{Prob:uplink}
& {\min_{{\bf{\tilde F}},{\bf{P}}}} \ \ \ \ {\rm{\overline{MSE}}}_{U}({\bf{\tilde F}},{\bf{P}}) \nonumber \\
& \ {\rm{s.t.}} \ \ \ \  {\rm{Tr}}({\bf{P}}_k{\bf{P}}_k^{\rm{H}}) \le P_{s,k}, \ \ k=1,\cdots,K \nonumber \\
& \ \ \ \ \ \ \ \ \ {\rm{Tr}}({\bf{\tilde F}}{\bf{\tilde
F}}^{\rm{H}}) \le P_{r} \nonumber \\
& \ \ \ \ \ \ \ \ \ {\bf{P}}={\rm{diag}}\{[{\bf{P}}_1,\ \cdots, \ {\bf{P}}_K]\}.
\end{align} Unfortunately, the optimization problem (\ref{Prob:uplink}) is still
nonconvex for ${\bf{\tilde F}}$ and ${\bf{P}}$, and thus there
is no closed-form solution. However, notice that if either
${\bf{\tilde F}}$ or ${\bf{P}}$ is fixed, the optimization
problem is convex with respect to the remaining variable. Therefore, an
iterative algorithm which designs ${\bf{\tilde F}}$ and
${\bf{P}}$ alternatively, is proposed as follows.

\noindent (1) \textbf{\underline{Design ${\bf{\tilde F}}$ when
${\bf{P}}$ is fixed}}

From (\ref{MSE_uplink_a}), it is noticed that ${\bf{\tilde F}}$ appears both inside and outside of the inverse operation. In order simplify the objective function, we use the following variant of matrix inversion lemma
\begin{align}
 {\bf{C}}^{\rm{H}}({\bf{C}}{\bf{C}}^{\rm{H}}+{\bf{D}})^{-1}{\bf{C}}&={\bf{I}}-({\bf{C}}^{\rm{H}}{\bf{D}}^{-1}{\bf{C}}+{\bf{I}})^{-1}.
\end{align} Taking ${\bf{C}}={\bf{H}}_{RB}{\bf{\tilde
F}}$ and ${\bf{D}}={\bf{R}}_{{\boldsymbol{\xi}}}$, the MSE
(\ref{MSE_uplink_a}) can be reformulated as \citep{Xing10}
\begin{align}
\label{MSE_uplink_c} {\rm{MSE}}_U({\bf{\tilde F}},{\bf{P}})
=&{\rm{Tr}}(({\boldsymbol{\Xi}}^{-1/2}{\bf{R}}_{{\bf{n}}}^{-1/2}{\bf{H}}_{MR}{\bf{P}})({\boldsymbol{\Xi}}^{-1/2}
{\bf{R}}_{{\bf{n}}_r}^{-1/2}
{\bf{H}}_{MR}{\bf{P}})^{\rm{H}}\nonumber \\
&  \times ({\bf{\tilde
F}}^{\rm{H}}{\bf{H}}_{RB}^{\rm{H}}{\bf{R}}_{{\boldsymbol{\xi}}}^{-1}{\bf{H}}_{RB}{\bf{\tilde
F}}+{\bf{I}})^{-1})
+{\rm{Tr}}(({\bf{P}}^{\rm{H}}{\bf{H}}_{MR}^{\rm{H}}{\bf{R}}_{{\bf{n}} }^{-1}{\bf{H}}_{MR}{\bf{P}}+{\bf{I}})^{-1}).
\end{align} Now, ${\bf{\tilde F}}$ only appears inside the matrix inverse.
If ${\bf{P}}$ is fixed, the last term of (\ref{MSE_uplink_c})
is independent of ${\bf{\tilde F}}$, and the optimization problem
(\ref{Prob:uplink}) becomes
\begin{align}
\label{prob:F_uplink} & {{\min_{{\bf{\tilde F}}}}} \ \ \ \
{\rm{Tr}}(\underbrace{({\boldsymbol{\Xi}}^{-1/2}{\bf{R}}_{{\bf{n}}}^{-1/2}{\bf{H}}_{MR}{\bf{P}})({\boldsymbol{\Xi}}^{-1/2}
{\bf{R}}_{{\bf{n}}}^{-1/2}
{\bf{H}}_{MR}{\bf{P}})^{\rm{H}}}_{\triangleq {\boldsymbol \Theta}}({\bf{\tilde F}}^{\rm{H}}\underbrace{{\bf{H}}_{RB}^{\rm{H}}{\bf{R}}_{{\boldsymbol{\xi}}}^{-1}{\bf{H}}_{RB}}_{\triangleq {\bf{M}}}
{\bf{\tilde F}}+{\bf{I}})^{-1}) \nonumber \\
& {\rm{s.t.}} \ \ \ \ \  {\rm{Tr}}({\bf{\tilde F}}{\bf{\tilde
F}}^{\rm{H}}) \le P_{r}.
\end{align} Based on eigen-decomposition, $
{\boldsymbol \Theta}={\bf{U}}_{{\boldsymbol \Theta}}{\boldsymbol \Lambda}_{{\boldsymbol \Theta}}{\bf{U}}_{{\boldsymbol \Theta}}^{\rm{H}}$ and $
{\bf{M}}={\bf{U}}_{{\bf{M}}}{\boldsymbol
\Lambda}_{{\bf{M}}}{\bf{U}}_{{\bf{M}}}^{\rm{H}}$, and defining
\begin{align}
\label{Lambda_F}
{\boldsymbol \Lambda}_{\bf{\tilde F}}\triangleq{\bf{U}}_{{\bf{M}}}^{\rm{H}}{\bf{\tilde F}}{\bf{U}}_{{\boldsymbol \Theta}},
\end{align}
the optimization problem (\ref{prob:F_uplink}) can be simplified as
\begin{align}
\label{prob:F_uplink_a} & {{\min_{{\boldsymbol \Lambda}_{\bf{\tilde
F}}}}} \ \ \ \
{\rm{Tr}}({\boldsymbol \Lambda}_{{\boldsymbol \Theta}}({\boldsymbol \Lambda}_{\bf{\tilde F}}^{\rm{H}}{\boldsymbol \Lambda}_{\bf{M}}{\boldsymbol \Lambda}_{\bf{\tilde F}}+{\bf{I}})^{-1}) \nonumber \\
& {\rm{s.t.}} \ \ \ \ \  {\rm{Tr}}({\boldsymbol \Lambda}_{\bf{\tilde F}}{\boldsymbol \Lambda}_{\bf{\tilde
F}}^{\rm{H}}) \le P_{r}.
\end{align}
Without loss of generality, the diagonal elements of ${\boldsymbol
\Lambda}_{{\boldsymbol \Theta}}$ and ${\boldsymbol
\Lambda}_{{\bf{M}}}$ are assumed to be arranged in decreasing order.
The closed-form solution of (\ref{prob:F_uplink_a}) can be shown to
be \citep{Xing10}
\begin{align}
\label{3_FFF}
{\boldsymbol \Lambda}_{\bf{\tilde F}}= \left[ {\begin{array}{*{20}c}
   {\left[\left(\frac{1}{\sqrt{\mu_f}}{\boldsymbol{\tilde
\Lambda}}_{\bf{M}}^{-1/2}{{\boldsymbol{\tilde
\Lambda}}_{\boldsymbol{\Theta}}^{1/2}} -{{\boldsymbol{\tilde
\Lambda}}_{\bf{M}}^{-1}} \right)^{+}\right]^{1/2}} & {{\bf{0}}_{L,N_R-L}}  \\
   {{\bf{0}}_{N_R-L,L}} & {{\bf{0}}_{N_R-L,N_R-L}}  \\
\end{array}} \right],
\end{align}where ${\boldsymbol {\tilde \Lambda}}_{\boldsymbol \Theta}$ and ${\boldsymbol {\tilde \Lambda}}_{\bf{M}}$ are the $L \times L$ principal submatrices of
 ${\boldsymbol {\Lambda}}_{\boldsymbol \Theta}$ and ${\boldsymbol {\Lambda}}_{\bf{M}}$,
 respectively. The scalar $\mu_f$ is the Lagrange multiplier which makes
${\rm{Tr}}({\boldsymbol \Lambda}_{\bf{\tilde F}}{\boldsymbol
\Lambda}_{\bf{\tilde F}}^{\rm{H}})=P_r$ hold. Based on
(\ref{Lambda_F}) and (\ref{3_FFF}), the optimal ${\bf{\tilde F}}$ can be recovered as
\begin{align}
\label{F_uplink} {\bf{\tilde
F}}={\bf{U}}_{{\bf{M}},L}\left[\left(\frac{1}{\sqrt{\mu_f}}{\boldsymbol{\tilde
\Lambda}}_{\bf{M}}^{-1/2}{{\boldsymbol{\tilde \Lambda}}_{\boldsymbol \Theta}^{1/2}}
-{{\boldsymbol{\tilde \Lambda}}_{\bf{M}}^{-1}}
\right)^{+}\right]^{1/2}{\bf{U}}_{{\boldsymbol \Theta},L}^{\rm{H}},
\end{align}where ${\bf{U}}_{{\bf{M}},L}$ and ${\bf{U}}_{{\boldsymbol \Theta},L}$ are
the first $L$ columns of ${\bf{U}}_{{\bf{M}}}$ and ${\bf{U}}_{{\boldsymbol \Theta}}$,
respectively. Finally, the optimal ${\bf{F}}$ is given by ${\bf{F}}={\bf{\tilde F}}
{\boldsymbol
\Xi}^{-1/2}{\bf{R}}_{\bf{n}}^{-1/2}$.

\noindent (2) \textbf{\underline{Design ${\bf{P}}$ when
${\bf{\tilde F}}$ is fixed}}

Since ${\boldsymbol \Xi}$ in (\ref{MSE_uplink_a}) depends on ${\bf{P}}$, the MSE expression in (\ref{MSE_uplink_a}) is a complicated function of ${\bf{P}}$, direct optimization of ${\bf{P}}$ seems intractable. However, based on the property of trace operator
${\rm{Tr}}({\bf{D}}{\bf{C}})={\rm{Tr}}({\bf{C}}{\bf{D}})$, the total
MSE (\ref{MSE_uplink_a}) can be reformulated as \citep{Xing201077}
\begin{align}
\label{MSE_uplink_b}
&{\rm{\overline{MSE}}}_U({\bf{\tilde F}},{\bf{P}})\nonumber \\
=&{\rm{Tr}}({\bf{I}}_L)-{\rm{Tr}}(({\bf{H}}_{RB}{\bf{\tilde
F}})^{\rm{H}} ({\bf{H}}_{RB}{\bf{\tilde F}}
{\bf{\tilde F}}^{\rm{H}}{\bf{H}}_{RB}^{\rm{H}}+{\bf{R}}_{{\boldsymbol{\xi}}})^{-1}\nonumber
\\
& \ \ \ \ \ \ \ \ \ \ \ \ \ \ \ \ \ \ \ \times
 {\bf{H}}_{RB}{\bf{\tilde F}})({\boldsymbol{\Xi}}^{-1/2}\underbrace{{\bf{R}}_{{\bf{n}}}^{-1/2}{\bf{H}}_{MR}{\bf{P}}{\bf{P}}^{\rm{H}}
 {\bf{H}}_{MR}^{\rm{H}}{\bf{R}}_{{\bf{n}}}^{-1/2}}_{={\boldsymbol \Xi}-{\bf{I}}}
{\boldsymbol{\Xi}}^{-1/2})) \nonumber \\
=&{\rm{Tr}}({\bf{I}}_L)-{\rm{Tr}}(({\bf{H}}_{RB}{\bf{\tilde
F}})^{\rm{H}} ({\bf{H}}_{RB}{\bf{\tilde F}}
{\bf{\tilde F}}^{\rm{H}}{\bf{H}}_{RB}^{\rm{H}}+{\bf{R}}_{{\boldsymbol{\xi}}})^{-1}
 {\bf{H}}_{RB}{\bf{\tilde F}})({\bf{I}}_{N_R}-{\boldsymbol \Xi}^{-1})).
\end{align}
Substituting the definition of ${\boldsymbol \Xi}$ into
(\ref{MSE_uplink_b}), the MSE can be further rewritten as
\begin{align}
\label{MSE_aaaaa}
&{\rm{\overline{MSE}}}_U({\bf{\tilde F}},{\bf{P}})\nonumber \\
=&{\rm{Tr}}(\underbrace{({\bf{H}}_{RB}{\bf{\tilde
F}})^{\rm{H}}({\bf{H}}_{RB}{\bf{\tilde F}} {\bf{\tilde
F}}^{\rm{H}}{\bf{H}}_{RB}^{\rm{H}}+{\bf{R}}_{{\boldsymbol{\xi}}})^{-1}
({\bf{H}}_{RB}{\bf{\tilde F}})}_{\triangleq
{\boldsymbol{\Pi}}}
\nonumber \\
&
\ \ \ \ \ \ \ \ \ \ \ \ \ \ \ \ \ \ \ \times ({\bf{R}}_{{\bf{n}}}^{-1/2}{\bf{H}}_{MR}{\bf{P}}{\bf{P}}^{\rm{H}}
{\bf{H}}_{MR}^{\rm{H}}{\bf{R}}_{{\bf{n}}}^{-1/2}+{\bf{I}}_{N_R})^{-1})\nonumber \\
& + {\rm{Tr}}({\bf{I}}_L)-{\rm{Tr}}({({\bf{H}}_{RB}{\bf{\tilde
F}})^{\rm{H}}({\bf{H}}_{RB}{\bf{\tilde F}} {\bf{\tilde
F}}^{\rm{H}}{\bf{H}}_{RB}^{\rm{H}}+{\bf{R}}_{{\boldsymbol{\xi}}})^{-1}
({\bf{H}}_{RB}{\bf{\tilde F}})}),
\end{align} where ${\bf{P}}$ only appears inside of the inverse operation.
As the last two terms of (\ref{MSE_aaaaa}) are independent of ${\bf{P}}$, the
optimization problem for ${\bf{P}}$ is
\begin{align}
\label{Prob:Q} & {\min_{{\bf{P}}}} \ \ \ \
{\rm{Tr}}({\boldsymbol{\Pi}}({\bf{R}}_{{\bf{n}}}^{-1/2}{\bf{H}}_{MR}{\bf{P}}{\bf{P}}^{\rm{H}}
{\bf{H}}_{MR}^{\rm{H}}{\bf{R}}_{{\bf{n}}}^{-1/2}+{\bf{I}}_{N_R})^{-1}) \nonumber \\
& {\rm{s.t.}} \ \ \ \ \   {\rm{Tr}}({\bf{P}}_k{\bf{P}}_k^{\rm{H}}) \le P_{s,k} \ \ k=1,\cdots,\ K \nonumber \\
& \ \ \ \ \ \ \ \ \  {\bf{P}}={\rm{diag}}\{[{\bf{P}}_1,\ \cdots, \ {\bf{P}}_K]\}.
\end{align}

With the definitions of ${\bf{H}}_{MR}$ and ${\bf{P}}$,
\begin{align}
\label{Q_Defin}
 &{\bf{H}}_{MR}{\bf{P}}{\bf{P}}^{\rm{H}}{\bf{H}}_{MR}^{\rm{H}}
=\sum_{k=1}^K\{{\bf{H}}_{MR,k}\underbrace{{\bf{P}}_k{\bf{P}}_k^{\rm{H}}}_{\triangleq
{\bf{Q}}_k}{\bf{H}}_{MR,k}^{\rm{H}}\}.
\end{align} Putting (\ref{Q_Defin}) into (\ref{Prob:Q}), the optimization problem becomes
\begin{align}
\label{Prob:Q_a} & {\min_{{\bf{Q}}_k}} \ \ \ \
{\rm{Tr}}({\boldsymbol{\Pi}}({\bf{R}}_{{\bf{n}}}^{-1/2}\sum_{k=1}^{K}\{{\bf{H}}_{MR,k}{\bf{Q}}_k
{\bf{H}}_{MR,k}^{\rm{H}}\}{\bf{R}}_{{\bf{n}}}^{-1/2}+{\bf{I}}_{N_R})^{-1}) \nonumber \\
& {\rm{s.t.}} \ \ \ \ \   {\rm{Tr}}({\bf{Q}}_k) \le P_{s,k}, \ k=1,\cdots,K, \nonumber  \\
& \ \ \ \ \ \ \ \ \  {\bf{Q}}_k \succeq {\bf{0}}.
\end{align}Using the Schur-complement lemma \citep{Horn85}, the optimization problem
(\ref{Prob:Q_a}) can be further formulated as a standard SDP
optimization problem \citep{Xing201077}
\begin{align}
\label{P_k}
& {\min_{{\bf{X}},{\bf{Q}}_k}} \ \ \ \ {\rm{Tr}}({\bf{X}}) \nonumber \\
& \ {\rm{s.t.}}  \ \ \ \ \left[ {\begin{array}{*{20}c}
   {\bf{X}} & {\boldsymbol{\Pi}}^{1/2}  \\
   {\boldsymbol{\Pi}}^{1/2} & {{\bf{R}}_{{\bf{n}}}^{-1/2}\sum_k\{{\bf{H}}_{MR,k}{\bf{Q}}_k{\bf{H}}_{MR,k}^{\rm{H}}\}{\bf{R}}_{{\bf{n}}}^{-1/2}+{\bf{I}}_{N_R}}  \\
\end{array}} \right]\succeq {\bf{0}} \nonumber \\
& \ \ \ \ \ \ \ \ \ \  {\rm{Tr}}({\bf{Q}}_{k}) \le P_{s,k}, \ \
k=1,\cdots, K \nonumber \\
&  \ \ \ \ \ \ \ \ \ \ {\bf{Q}}_k \succeq {\bf{0}}.
\end{align} The SDP problems can be efficiently solved using interior-point polynomial
algorithms \citep{Boyd04}.

In summary, when $N_{M,k} \le L_k$, the uplink beamforming design
alternates between the design of ${\bf {\tilde F}}$ in
(\ref{F_uplink}) and ${\bf{Q}}_k$ in (\ref{P_k}). The algorithm
stops when $\|{\rm{MSE}}_U^{I}-{\rm{MSE}}_U^{I+1}\| \le
{\mathcal{T}}_U$, where ${\rm{MSE}}_U^{I}$ is the total MSE in the
  $I^{\rm{th}}$ iteration and $\mathcal{T}_U$ is a threshold value.
After convergence, ${\bf{P}}_k={\bf{Q}}_k^{1/2}$,
${\bf{F}}={\bf{\tilde F}} {\boldsymbol
\Xi}^{-1/2}{\bf{R}}_{\bf{n}}^{-1/2}$ and $
{\bf{B}}=({\bf{H}}_{RB}{\bf{F}}{\bf{H}}_{MR}{\bf{P}})^{\rm{H}}({\bf{H}}_{RB}{\bf{F}}
{\bf{R}}_{{\bf{x}}}{\bf{F}}^{\rm{H}}{\bf{H}}_{RB}^{\rm{H}}+
{\bf{R}}_{{\boldsymbol{\xi}}})^{-1}$. We refer the algorithm in this section as Algorithm 2.

\noindent {\textbf{Remark 1}}: In case $N_{M,k} >L_k$, there is an
additional constraint ${\rm{Rank}}\{ {\bf{Q}}_k \} \le N_k $ in
(\ref{Prob:Q_a}). In this case, as rank constraints are nonconvex,
transition from (\ref{Prob:Q_a}) to (\ref{P_k}) involves a
relaxation on the rank constraint. Then the objective function of
(\ref{P_k}) is a lower bound of that of (\ref{Prob:Q_a}). However,
this problem seems to be common to all multiuser MIMO uplink
beamforming \citep{Codreanu07}, \citep{Hunger09}. Notice that when
$N_{M,k} \le L_k$, there is no relaxation involved.

\subsection{Special cases}

Notice that (\ref{F_uplink}) has a more general form than the water-filling solution in traditional point-to-point MIMO systems. On the other hand, (\ref{P_k}) is a SDP problem frequently encountered in multiuser MIMO systems. In particular, they include the following existing
algorithms as special cases.

\noindent $\bullet$ If ${\bf{H}}_{RB}={\bf{I}}_{L}$ and
${\bf{R}}_{\boldsymbol \xi}={\bf{0}}_{L,L}$, we have
${\boldsymbol{\Pi}}={\bf{I}}_{L}$
 in (\ref{Prob:Q_a}), and the SDP optimization problem (\ref{P_k}) reduces to that of the uplink multiuser MIMO systems
\citep{Codreanu07}, \citep{Hunger09}. Therefore, they have the same
solution.

\noindent $\bullet$ Substituting $K=1$ and ${\bf{P}}={\bf{I}}_{L_1}$
into (\ref{F_uplink}), it reduces to the solution proposed for LMMSE
joint design of relay forwarding matrix and destination equalizer in
AF MIMO relay systems  without source precoder \citep{Guan08}.

\noindent $\bullet$ Notice that when there is only one mobile
terminal ($K=1$), the optimization problem (\ref{Prob:Q}) is in the
same form as (\ref{prob:F_uplink}). Defining
${\bf{H}}_{MR}^{\rm{H}}{\bf{R}}_{{\bf{n}}}^{-1}{\bf{H}}_{MR}={\bf{U}}_{MR}{\boldsymbol
{\Lambda}}_{MR}{\bf{U}}_{MR}^{\rm{H}}$, and ${\boldsymbol
\Pi}={\bf{U}}_{\boldsymbol \Pi}{\boldsymbol {\Lambda}}_{\boldsymbol
\Pi}{\bf{U}}_{\boldsymbol \Pi}^{\rm{H}}$, a closed-form solution can
be derived using the same procedure as for ${\bf{\tilde F}}$, and we
have
\begin{align}
\label{P_uplink}
{\bf{P}}={\bf{U}}_{MR,L}\left[\left(\frac{1}{\sqrt{\mu_p}}{\boldsymbol{\tilde
\Lambda}}_{MR}^{-1/2}{{\boldsymbol{\tilde \Lambda}}_{\boldsymbol
\Pi}^{1/2}} -{{\boldsymbol{\tilde \Lambda}}_{MR}^{-1}}
\right)^{+}\right]^{1/2}
\end{align} where the ${\boldsymbol {\tilde \Lambda}}_{MR}$ and
${\boldsymbol {\tilde \Lambda}}_{\boldsymbol \Pi}$ are the $L \times L$
principal submatrices of
 ${\boldsymbol {\Lambda}}_{MR}$ and ${\boldsymbol {\Lambda}}_{\boldsymbol \Pi}$,
 respectively, and the matrix ${\bf{U}}_{MR,L}$ is the
first $L$ columns of ${\bf{U}}_{MR}$. The scalar $\mu_p$ is the
Lagrange multiplier which makes
${\rm{Tr}}({\bf{P}}{\bf{P}}^{\rm{H}})=P_{s,1}$ hold. In this case,
the solution given by (\ref{P_uplink}) corresponds to the source
precoder design  for  AF MIMO relay systems with single user
\citep{Rong09}.

\noindent $\bullet$ Furthermore, substituting
${\bf{H}}_{RB}={\bf{I}}_{L}$ and ${\bf{R}}_{\boldsymbol
\xi}={\bf{0}}_{L,L}$ into (\ref{P_uplink}), it becomes the
closed-form solution for LMMSE transceiver design in point-to-point
MIMO systems \citep{Sampath01}.

\section{Simulation Results and Discussions}
\label{sect:simul}

In this section, we investigate the performance of the proposed
algorithms for downlink and uplink. In the simulations, there is one
BS, one relay station and two mobile terminals. For each mobile
terminal, two independent data streams will be transmitted in the
uplink (or received in the downlink) simultaneously. For each data
stream, 10000 independent QPSK symbols are transmitted. The elements
of MIMO channels between BS and relay station and between relay
station and mobile terminals are generated as independent complex
Gaussian random variables with zero mean and unit variance.  Each
point in the following figures is an average of 500 independent
channel realizations. In order to solve SDP problems, the widely
used optimization matlab toolbox CVX is adopted \citep{Grant07}. The
thresholds for terminating the iterative algorithms are set at
${\mathcal{T}}_D={\mathcal{T}}_U=0.0001$.

\begin{figure}[!ht]
\centering
\includegraphics[width=.6\textwidth]{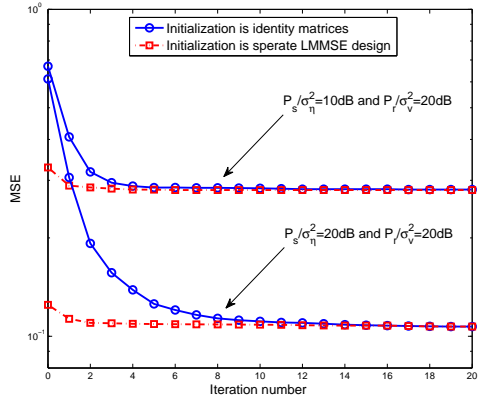}
\caption{The convergence behavior of the proposed Algorithm 1 when $N_B=4$, $N_R=4$ and $N_{M,k}=2$ with 2 users.}\label{fig:3}
\end{figure}

First, let us focus on the downlink. In downlink, the noise covariance matrices at relay station and mobile terminals are
${\bf{R}}_{{\boldsymbol{\eta}}}=\sigma_{\eta}^2{\bf{I}}_{N_R}$ and
${\bf{R}}_{{\bf{v}}_{1}}={\bf{R}}_{{\bf{v}}_{2}}=\sigma_{v}^2{\bf{I}}_{N_M}$, respectively.
We define the first hop SNR at the relay station as
$P_s/{\sigma_{\eta}}^2$, and the second hop SNR at mobile
terminals as $P_r/\sigma_v^2$.
Fig.~\ref{fig:3} shows the convergence behavior of the proposed Algorithm 1 for downlink with different second hop SNR at mobile terminals when $N_B=4$, $N_R=4$, $N_{M,k}=2$. Both initializations with identity matrices and the separate LMMSE design are shown. It can be seen that the proposed algorithm converges quickly, within 20 iterations. Furthermore, the convergence speed with separate LMMSE design as initialization is faster than that with identity matrices. It can also be seen that the two initializations result in the same MSE after convergence.

\begin{figure}[!ht]
\centering
\includegraphics[width=.6\textwidth]{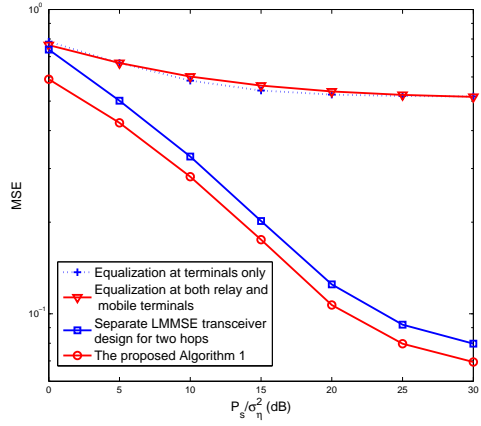}
\caption{Total MSEs of detected data of the proposed
Algorithm 1 and suboptimal algorithms, when $N_B=4$,
$N_R=4$, $N_{M,k}=2$ and $P_r/\sigma_{v}^2$=20dB.}\label{fig:4}
\end{figure}

Fig.~\ref{fig:4} compares the total data MSEs of the proposed
Algorithm 1 and several suboptimal algorithms versus the first hop SNR $P_s/\sigma_{\eta}^2$.  The second hop SNR at mobile terminals is fixed to be 20dB. The number of antennas is set as $N_B=4$, $N_R=4$ and $N_{M,k}=2$. The suboptimal algorithms under consideration are

\noindent $\bullet$ Direct amplify-and-forward, in which the
precoder ${\bf{T}}$ at BS and forwarding matrix ${\bf{W}}$ at relay
are proportional to identity matrices.  At mobile terminals, LMMSE
equalizer for the combined first hop and second hop channel is
adopted to recover the signal \citep{Guan08}.

\noindent $\bullet$ The first hop channel is equalized at relay and then the second hop channel is equalized at mobile terminals, both with LMMSE equalizers.

\noindent $\bullet$ Separate LMMSE design proposed for initialization of Algorithm 1.

\noindent From Fig.~\ref{fig:4}, it can be seen that as there is no
precoder design at BS for the first two suboptimal algorithms, the data streams at different terminals
cannot be efficiently separated by linear equalizers, resulting in poor performances. The separate LMMSE transceiver design has a much better performance. On the other hand, the proposed Algorithm 1 has the best performance among the four algorithms. The gap between the MSEs of the separate LMMSE design and that of Algorithm 1 is the performance gain obtained by additional iterations.

\begin{figure}[!ht]
\centering
\includegraphics[width=.6\textwidth]{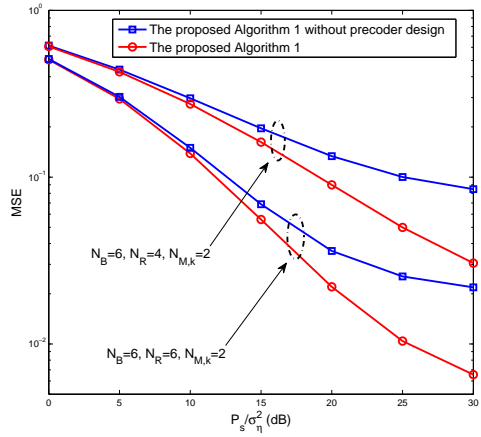}
\caption{Total MSEs of detected data of the proposed
Algorithm 1 with and without precoder
design.}\label{fig:5}
\end{figure}

As the proposed Algorithm 1 involves a computational expensive SDP
for the precoder ${\bf{T}}$ design, it is of great interest to investigate
how much degradation would result from skipping the precoder design.
Fig.~\ref{fig:5} compares the total data MSEs of the proposed
Algorithm 1 and the same algorithm but fixing the precoder
${\bf{T}} \propto {\bf{I}}$. The second hop SNR at mobile terminals
$P_r/\sigma_v^2$ is fixed to be 20dB. From Fig.~\ref{fig:5}, it can
be seen that a properly designed precoder significantly improves the system
performance when the first hop SNR is high.  Without the precoder,
the data MSEs exhibit error floors at much lower $P_s/\sigma_{\eta}^2$. On the other hand, we can also see that
increasing the number of antennas at the relay station greatly
improves the system performance, as it simultaneously increases the
diversity gain of the two hops.

Now, let us turn to the results in the uplink. In uplink case, the
noise covariance matrices at relay station and BS are
${\bf{R}}_{{\bf{n}}}=\sigma_n^2{\bf{I}}_{N_R}$ and
${\bf{R}}_{{\boldsymbol{\xi}}}=\sigma_{\xi}^2{\bf{I}}_{N_B}$,
respectively. We define the fist hop SNR at the relay station as
$P_{s}/{\sigma_{n}}^{2}$, where $P_s=\sum_{k=1}^K P_{s,k}$. The
second hop SNR at the BS is defined as $P_r/\sigma_{\xi}^2$.

\begin{figure}[!ht]
\centering
\includegraphics[width=.6\textwidth]{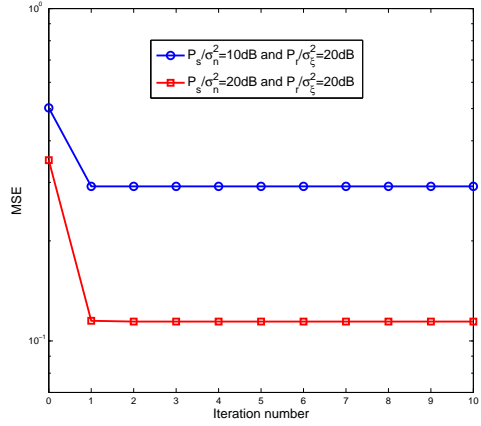}
\caption{The convergence behavior of Algorithm 2 for
uplink when $N_B=4$, $N_R=4$ and $N_{M,k}=2$.}\label{fig:6}
\end{figure}

Fig.~\ref{fig:6} shows the convergence behavior of the proposed
Algorithm 2 for uplink when $N_B=4$, $N_R=4$ and $N_{M,k}=2$. Notice
that in this case, at each mobile terminal the number of antennas
equals to that of the data streams, and Algorithm 2 involves no
relaxation. The initialization is identity matrices. It can be seen
that Algorithm 2 converges very fast, indicating its superior
performance.

\begin{figure}[!ht]
\centering
\includegraphics[width=.64\textwidth]{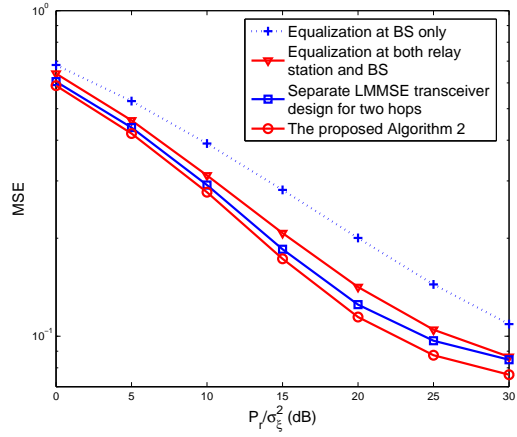}
\caption{Total MSEs of detected data of Algorithm 2 and
suboptimal algorithms, when $N_B=4$, $N_R=4$, $N_{M,k}=2$ and
$P_{s}/\sigma_{n}^2$=20dB.}\label{fig:7}
\end{figure}

Fig.~\ref{fig:7} shows the total data MSEs of the proposed Algorithm
2 and suboptimal algorithms, when $N_B=4$, $N_R=4$, $N_{M,k}=2$ and
the SNR at relay station $P_s/\sigma_{n}^2$ is fixed to be 20dB. The
suboptimal algorithms are similar to those for the downlink. In
particular, we consider

\noindent $\bullet$ Equalization of the equivalent two-hop channel is applied only at the BS.

\noindent $\bullet$ Equalization is applied at relay station for the mobile-to-relay channel, and also at BS for the relay-to-BS channel.

\noindent $\bullet$ Separate LMMSE design. The first hop is
considered as a traditional multiuser MIMO uplink system, and the
beamforming matrices are designed using the algorithms in
\citep{Shi08} and \citep{Codreanu07}. The second hop is considered
as a point-to-point MIMO system, and the beamforming matrices are
designed using the result in \citep{Sampath01}.

\noindent  From Fig.~\ref{fig:7}, it can be seen that the
performance of the proposed Algorithm 2 is better than other
suboptimal algorithms. However, as the signals from different
terminals are cooperatively detected at BS, the gaps between the
performance of the suboptimal algorithms from that of Algorithm 2 is
much smaller compared to their counterparts in downlink.

\begin{figure}[!ht]
\centering
\includegraphics[width=.64\textwidth]{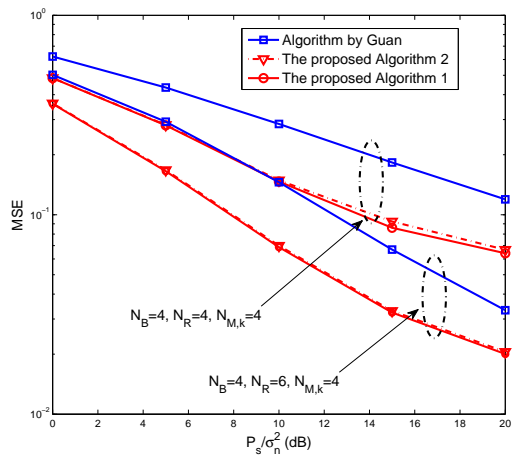}
\caption{Total MSEs of the detected data of the
 Algorithm 1, Algorithm 2 with relaxation and the algorithm proposed in \citep{Guan08}.}\label{fig:8}
\end{figure}

When $L_k<N_{M,k}$ in the uplink, strictly speaking, Algorithm 2
involves a relaxation, and its performance is not guaranteed.
However, a simple variation of Algorithm 1 can be used for
beamforming design in this case. Fig.~\ref{fig:8} shows the total
data MSEs of Algorithm 1 for uplink and Algorithm 2 with rank
relaxation, when $L_k=2$ and $N_{M,k}=4$. The SNR at BS is fixed at
$P_r/\sigma_{\xi}^2$=20dB. The joint relay forwarding matrix and
destination equalizer design in \citep{Guan08} is also shown for
comparison. It can be viewed as a design without source precoders at
mobile terminals. From Fig.~\ref{fig:8}, it can be seen that
Algorithm 1 and Algorithm 2, which involve the joint design of
precoder, forwarding matrix and equalizer perform better than the
algorithm in \citep{Guan08}. This indicates the importance of source
precoder design in AF relay cellular networks. Furthermore, although
Algorithm 2 involves a relaxation, its performance is still
satisfactory, and is close to that of Algorithm 1. Finally, it can
also be concluded that increasing the number of antennas at relay
station can greatly improve the performance of uplink beamforming
design for all algorithms.

\section{Conclusions}
\label{sect:conclusion} In this paper, LMMSE beamforming design for
amplify-and-forward MIMO relay cellular networks has been investigated.
Both uplink and downlink cases were considered. In the downlink,
precoder at base station, forwarding matrix at relay station and
equalizer at mobile terminals were jointly designed by an iterative algorithm. On the other
hand, in the uplink case, we demonstrated that in general the beamforming design problem can be solved by an iterative algorithm with the same structure as in the downlink case.  Furthermore, for the fully loaded or overloaded uplink systems, a novel beamforming design algorithm was derived and it includes several existing algorithms for conventional point-to-point or multiuser systems as special cases. Finally, simulation results were presented to show the performance advantage of the proposed
algorithms over several suboptimal schemes.

\end{document}